# Renormalization group approach to power-law modeling of complex metabolic networks

Benito Hernández-Bermejo

*Departamento de Física, Universidad Rey Juan Carlos.*

*Calle Tulipán S/N, 28933-Móstoles-Madrid. Spain.*

Phone: (+34) 914 88 73 91. Fax: (+34) 916 64 74 55.

E-mail: benito.hernandez@urjc.es

**Abstract**

In the modeling of complex biological systems, and specially in the framework of the description of metabolic pathways, the use of power-law models (such as S-systems and GMA systems) often provides a remarkable accuracy over several orders of magnitude in concentrations, an unusually broad range not fully understood at present. In order to provide additional insight in this sense, this article is devoted to the renormalization group analysis of reactions in fractal or self-similar media. In particular, the renormalization group methodology is applied to the investigation of how rate-laws describing such reactions are transformed when the geometric scale is changed. The precise purpose of such analysis is to investigate whether or not power-law rate-laws present some remarkable features accounting for the successes of power-law modeling. As we shall see, according to the renormalization group point of view the answer is positive, as far as power-laws are the critical solutions of the renormalization group transformation, namely power-law rate-laws are the renormalization group invariant solutions. Moreover, it is shown that these results also imply invariance under the group of concentration scalings, thus accounting for the reported power-law model accuracy over several orders of magnitude in metabolite concentrations.

# 1. Introduction

## 1.1. Power-law modeling

Power-law models constitute a highly structured nonlinear representation of many complex biological systems. Such models, originally developed on the basis of a first order Taylor series in log-log space, have provided a very powerful modeling framework, specially (but not only) in the domain of biochemical pathways. For instance, see Savageau (1976); Voit (2000); Voit *et al.* (1991) for a general and detailed description of the formalism and its applications. As indicated, one important advantage of such formalism is that the mathematical description of complex systems is highly structured, thus allowing a remarkable mathematical tractability, let it be in the form of S-systems

$$\frac{dx_i}{dt} = a_i \prod_{j=1}^{n} x_j^{g_{ij}} - b_i \prod_{j=1}^{n} x_j^{h_{ij}}, \qquad i = 1,\ldots,n$$

or of generalized mass-action (or GMA, in what follows) systems:

$$\frac{dx_i}{dt} = \sum_{j=1}^{p} a_{ij} \prod_{k=1}^{n} x_k^{g_{ijk}} - \sum_{j=1}^{q} b_{ij} \prod_{k=1}^{n} x_k^{h_{ijk}}, \qquad i = 1,\ldots,n$$

In both cases, and in the biochemical context, the system variables $x_i$ describe metabolite concentrations. Therefore, both S-systems and GMA systems are purely reactional models, the power-law functions describing the kinetic rate-laws (also termed velocity functions) of the reactions being modeled. An additional and very significant feature of modeling based on either S-systems or GMA systems is that, typically, such models are valid over several orders of magnitude in concentrations. This behavior has been widely reported in the literature (Savageau, 1976; Sorribas and Savageau 1989a, 1989b, 1989c; Voit and Savageau, 1987; Voit *et al.* 1991). In fact, it is not uncommon that such models provide accurate predictions over variations of two or three orders of magnitude in concentrations, namely it is possible to find sizes as large as 100- or 1000-fold for the range of validity of the models. In spite that power-law models were in principle conceived as precise nonlinear approximations, such success seems to suggest that more fundamental reasons might exist in such a way that power-law models could, in fact, be the natural description for many processes, specially in a biochemical framework. In the last context, an equivalent question is: to what extent power-law rate-laws do implement some fundamental properties of reactions, that could explain the accuracy of power-law models? In the literature, different approaches and explanations have been proposed in order to account for such features. In this sense, we can mainly mention the mathematical nature of the formalism, the existence of systemic regulations that maintain the concentrations within narrow limits, or the advantages of the agregation of interactions in the case of S-systems (Voit *et al.* 1991). Other arguments are derived from considerations based on approximation theory (Savageau, 1979a, 1979b) as well as on the role of fractal kinetics (Aon *et al.* 2004; Kopelman, 1986, 1991; Savageau, 1993, 1995, 1998). In particular, it is interesting to recall the important role of fractals in the context of *in vivo* chemistry, as far as many reactions are confined to two-dimensional membranes, one-dimensional channels, or fractal domains of non-integer dimension, and an alternative to the classical framework is necessary for the analysis of such phenomena. All the contributions just mentioned provide very valuable insights that help understanding the naturalness of the power-law function as a tool for modeling, in spite that the ultimate reasons accounting for this success seem to be not yet understood. In this context, the methods of group theory in general (and of the powerful tools of renormalization group in particular) have not been exploited in order to analyze the fundamental reasons that confer on power-laws such a special role. Actually, the different studies about power-law systems based on group-theoretic properties (Díaz-Sierra *et al.* 1999; Hernández-Bermejo and Fairén, 1997; Voit, 1992) are not devoted to the reported accuracy of power-law models. In fact, such papers deal mainly with mathematical properties of power-law systems, such as the existence of first integrals or reduction procedures. Precisely, the purpose of the present work is to make use of renormalization group methods in order to shed some light on the fundamental properties accounting for the success of power-law functions in the modeling through several orders of magnitude in concentrations. In first place, let us recall some general features regarding the renormalization group in order to provide a self-contained presentation.

*1.2. Brief outline of the renormalization group methods*

Renormalization group techniques constitute a significant tool of physics. Actually, renormalization group methods are very diverse, ranging from quantum field theory to statistical mechanics. The last perspective, naturally associated to real space transformations, will be the one of interest in what follows. For instance, see Creswick *et al.* (1992); Fisher (1998); Schroeder (1991); Takayasu (1990) for some clear introductions to renormalization group methods in statistical mechanics, including applications to fractals and self-similarity.

The purpose of the renormalization group is to treat quantitatively the change of a physical magnitude when the geometric scale is changed. Here the term "physical" is to be understood in a broad sense. For instance, let $q$ be a certain physical quantity measured at a certain scale of coarse-graining, and let $q^*$ be the same quantity measured with the scale of coarse-graining changed by a factor $\lambda$. The rescaled value $q^*$ can be related to the original value of $q$ by an appropriate transformation $q^* = f_\lambda(q)$. This transformation $f_\lambda$ is the renormalization group transformation for magnitude $q$. Now the basis of the renormalization group analysis proceeds to the search of a fixed point of the renormalization group transformation, namely the search of a critical value $q_c$ such that $q_c = f_\lambda(q_c)$. The value $q_c$ thus corresponds to a renormalization group invariant solution, as far as it remains unaltered after the application of the renormalization group transformation. In statistical mechanics, such solutions usually correspond to critical values of different processes such as phase transitions, percolation, etc. The reason is clear: after successive application of the renormalization group transformation, we find $q_c = f_\lambda(f_\lambda(q_c))$. This explains also the close relationship of the renormalization group invariant solutions and fractality. In fact, historically the notions of fractal and renormalization group appeared independently, but both were intended to analyze what is invariant under the change of scale of observation: fractal for geometrical objects, and renormalization group for physical quantities. It is also important to stress that, except for very simple problems, the renormalization group transformation is generally an approximation. In other words, usually $f_\lambda$ is not the exact description of how $q$ changes after the scaling. Instead, it is in general a simplified approximation in order to (i) have the possibility of explicitly writing a renormalization group transformation $f_\lambda$, and (ii) find a transformation $f_\lambda$ that is amenable to analysis. This implies also that the solutions found by means of renormalization group techniques often provide a hint about the system properties, usually qualitative (and only approximate from a quantitative point of view) rather that leading to a precise analysis. Of course, the previous considerations imply also that a given problem can be analyzed by means of the renormalization group methods with progressively refined degrees of approximation.

In a biological context, the origin of power-law or allometric relationships has deserved a significant interest from many decades (e.g. see Huxley, 1932). In this sense, very diverse contributions can be found in the literature, for instance those based on statistical approaches (Kaitaniemi, 2004; Packard, 2009; Wu *et al.*, 2002), network theory (Furusawa and Kaneko, 2006), and on physical arguments, mainly based on fractality and scaling principles (see Aon *et al.* 2004; Auffray and Nottale, 2008; Demetrius, 2006; West, 1999; West *et al.* 2002; West and Brown, 2005). In addition, some authors have specifically addressed the relevance of group theory and the renormalization group approach (Derome, 1977; West, 2004) in such framework. However, as indicated in the previous subsection, the renormalization group methodology has not yet been applied in the context of power-law models (such as S-systems and GMA systems) and power-law rate-laws in order to account for the accuracy of the resulting models. As indicated, this is just the purpose of the present work. More precisely, in this article the renormalization group analysis of reactions in fractal or self-similar media is considered. In particular, the renormalization group methodology is applied to the investigation of how rate-laws describing such reactions are transformed when the geometric scale is changed. This will be done in two successive steps corresponding to different degrees of refinement in the renormalization group approximation. Such steps shall be termed Zero-Order renormalization group approximation (Section 2) and First-Order renormalization group approximation (Section 3). The precise purpose of such analysis is to investigate whether or not power-law rate-laws present some remarkable features accounting for the successes of power-law modeling. As we shall see, according to the renormalization group point of view the answer is positive, as far as power-laws are the critical solutions of the renormalization group transformation, namely power-law rate-laws are the renormalization group invariant solutions. Moreover, it is shown that these results also imply the system invariance under the group of concentration scalings, thus accounting for the reported power-law model accuracy over several orders of magnitude in metabolite concentrations. The work concludes in Section 4 with a discussion of the results found.

## 2. Zero-Order renormalization group approximation: exact solution

*2.1. One-variable case*

Let us first consider the simplest situation in which a single metabolite $A$ reacts in a given region of volume $V$ in which the reacting species is restricted to a region of fractal dimension $d$. In the following, we shall denote by $M$ the initial number of molecules in that volume, and therefore the concentration of the metabolite is $x=M/V$. In addition, let $N$ be the increment in the number of molecules of $A$ in that region per time unit due to the reaction. Thus, $N>0$ implies a net production of $A$, while $N<0$ means that $A$ is depleted. Accordingly, the rate of the reaction is given by $v=N/V$. Assume now that due to the underlying fractality, the system is self-similar after a scaling, and let $\lambda$ be the scaling factor, namely we postulate the invariance of the system in a scale $\lambda$ times larger, and let us denote with a star superscript the system parameters in the new scale. Obviously, now we have $V^*=\lambda^3 V$ but notice that it is $M^*=\lambda^d M$. Consequently, for the concentration we find $x^*=M^*/V^*=\lambda^{d-3} x$. Now, it is postulated that the invariance of the system in the new scale can be quantitatively expressed as:

$$\frac{N}{MV} = \frac{N^*}{M^* V^*} \qquad (1)$$

In other words, the system invariance is assumed to mean that the fraction of molecules reacting per unit of time and per unit of volume remains invariant after the scaling. Thus, after (1) we readily find that $N^*=NM^*V^*M^{-1}V^{-1}=\lambda^{d+3} N$, as well as $v^*=N^*/V^* = \lambda^d v$. Since the reactional rate-law $v(x)$ is actually a function of the concentration, the scaling condition $v^*=\lambda^d v$ must be explicitly written as an identity between functions, namely $v^*(\xi)=\lambda^d v(\xi)$ for any argument $\xi$. This identity for $v$ is specially important for what is to follow, as far as it will be necessary for the application of the renormalization group condition. In fact, our main goal is to characterize the renormalization group invariant rate-laws. As indicated in the Introduction, the renormalization group condition assumes the scale invariance of the system, in our case in the context of the invariance of the rate-law. Consequently, in our case the renormalization group condition amounts to $v^*(x^*) = v(x)$. This identity can be combined with the scaling condition $v^*(\xi) = \lambda^d v(\xi)$ in order to explore the implications of renormalization group invariance in several cases.

To begin with, let us consider the case of a power-law functional form: $v(x) = v_0 x^g$. We thus have for the renormalization group condition: $v^*(x^*) = \lambda^d v(\lambda^{d-3} x) = \lambda^{d+g(d-3)} v_0 x^g = v_0 x^g = v(x)$. Then, after equating both functions, it is found that the renormalization group condition amounts to: $\lambda^{d+g(d-3)} = 1$. This equation admits a solution that actually provides invariance for all scales (i.e. for all values of $\lambda$). Such solution relates the renormalization group value of the kinetic order $g$ with the fractal dimension $d$ of the region. Thus the renormalization group value is:

$$g = \frac{d}{3-d} \quad , \quad 0 < d < 3 \tag{2}$$

This result presents several interesting features. First of all, notice that $g$ is positive, but can take arbitrarily large values. It is worth emphasizing that the renormalization group method is an approximation. Let us ignore for the moment the formal divergence at $d=3$, which is a consequence of the approximate character of the renormalization group approach in the Zero-Order case (note that for $d=3$ the only solution corresponds to the transformation for which $\lambda=1$). Such divergence will dissapear in the First-Order renormalization group approach to be presented later. The important feature at this stage is that the kinetic order increases when the fractal dimension grows, and such kinetic order can take a wide range of possible values. This is a reasonable property since the dimensional restriction of the medium should limit the ability of the metabolite to react. In fact, we see that $g$ tends to zero when $d$ also tends to zero, which is to be expected. The most important conclusion here is that, in this approximation, the power-law rate-law is an exact solution of the renormalization group condition for all values of $\lambda$.

Let us now compare this result with other possible functional forms. For instance, let us first turn to kinetic functions of the kind $v(x) = v_0 x^g / (K + x^g)$, with $g>0$. This form includes as particular cases the well-known Michaelis-Menten ($g=1$) and Hill rate-laws. Once the renormalization group condition $v^*(x^*) = v(x)$ is imposed, we readily find:

$$\frac{1}{K+x^g} = \frac{\lambda^{d+g(d-3)}}{K+\lambda^{g(d-3)}x^g}$$

Clearly, this outcome does not allow the renormalization group invariance in general. Anyway, such invariance can be approximately found in the limit case $K >> x^g$. In this situation, i.e. if $x^g$ can be considered negligible when compared to $K$ (for instance, in the case of sufficiently low concentration) then the kinetic function can be approximated by a power-law: $v(x) \approx (v_0/K)x^g$. Of course, this functional dependence was already considered before. Accordingly, kinetic functions of the form $v(x)=v_0 x^g/(K+x^g)$ do not satisfy the renormalization group invariance condition.

As an additional possibility, we shall consider the case of quasi-polynomial kinetic functions, namely: $v(x) = \sum_{i=1}^{q} a_i x^{g_i}$, where the $g_i$ are in general real numbers. We thus impose the renormalization group invariance condition $v^*(x^*) = v(x)$ and the outcome is that the following identity must be verified:

$$\sum_{i=1}^{q} a_i x^{g_i} = \lambda^d \left( \sum_{i=1}^{q} a_i \lambda^{g_i(d-3)} x^{g_i} \right)$$

Consequently, renormalization group invariance exists if and only if $d+g_i(d-3)=0$ for every $i=1,...,q$. In turn, this implies that $g_1 = g_2 = ... = g_q = d/(3-d)$, namely all exponents are equal and then the quasi-polynomial is actually a power-law. Otherwise, renormalization group invariance is not present for any quasi-polynomial not being a power-law.

We thus conclude that the power-law seems to be the only evident renormalization group invariant kinetic function in the single-variable case. Clearly, the same property should remain essentially valid in the generalization to several reacting metabolites. The investigation of this issue is the aim of the next subsection.

*2.2. Generalization to the n-variable case*

Let us now consider the generalized situation in which several metabolites $A_1,...,A_n$ react in a region of volume $V$ and fractal dimension $d$. We denote by $M_1,...,M_n$ the initial number of molecules of each kind in that volume, the concentrations of the metabolites thus being $x_i=M_i/V$. As before, $N_i$ will describe the variation in the number of molecules of $A_i$ that are present in that region per unit of time due to the reactions taking place, and then the reaction velocities are $v_i=N_i/V$. Let us assume again that the system is self-similar after a scaling of factor $\lambda$, and denote with a star superscript the system parameters in the new scale. Now we have $V^*=\lambda^3 V$ but $M_i^*=\lambda^d M_i$. Then, for the concentrations we find $x_i^*=M_i^*/V^*=\lambda^{d-3} x_i$. As we did in the previous subsection, it is postulated that the dynamical invariance of the system in the new scale can be quantitatively expressed as:

$$\frac{N_i}{M_i V} = \frac{N_i^*}{M_i^* V^*} \qquad (3)$$

for all $i=1,...,n$. In other words, now the system invariance is assumed to mean that the fraction of molecules of each species reacting per unit of time and per unit of volume remains invariant after the scaling. Thus, after (3) we again find that $N_i^*=\lambda^{d+3} N_i$, as well as $v_i^*=N_i^*/V^* = \lambda^d v_i$. Recalling that the rate-law $v_i(x_1,...,x_n)$ is actually described as a function of the concentrations, the scaling condition $v_i^*=\lambda^d v_i$ must be understood as an identity between functions, namely $v_i^*(\xi_1,...,\xi_n)= \lambda^d v_i(\xi_1,...,\xi_n)$ for any $n$-variable argument $(\xi_1,...,\xi_n)$. Again, our main goal is to characterize the renormalization group invariant rate-laws. Following a reasoning similar to the one for the single variable case, the renormalization group invariance criterion now amounts to $v_i^*(x_1^*,...,x_n^*) = v_i(x_1,...,x_n)$ for all $i=1,...,n$. Let us now analyze the implications of this condition for several functional dependences.

In first place, we consider the case of power-law functions:

$$v_i(x_1,\ldots,x_n) = v_{i0} \prod_{j=1}^{n} x_j^{g_{ij}} \quad , \quad i=1,\ldots,n$$

After some algebra, the joint application of the renormalization group invariance criterion and the scaling condition leads to the following compatibility equations:

$$\sum_{j=1}^{n} g_{ij} = \Gamma_i = \frac{d}{3-d} \quad , \quad 0 < d < 3, \quad i=1,\ldots,n \tag{4}$$

For future convenience, we shall define $\Gamma \equiv \Gamma_1 = \ldots = \Gamma_n$. Obviously, equation (4) is a generalization of equation (2). As it was the case in the single-metabolite situation, equation (4) constitutes a solution that actually provides an exact invariance for all scales (i.e. for all values of $\lambda$). This result presents several interesting features additional to those found for a single variable. First of all, note that condition (4) admits an infinity of different compatible solutions for the kinetic orders $g_{ij}$, including the possibility of simultaneous negative and positive values. Again, the average kinetic order increases when the fractal dimension grows, showing that the dimensional restriction of the medium should limit the ability of the metabolites to react. Once again, the most relevant conclusion is that, in this approximation, the power-law rate-law is an exact solution of the renormalization group condition for all values of $\lambda$ (perfect self-similarity).

Due to the fact that equation (4) admits an infinity of solutions for the kinetic orders $g_{ij}$, now the renormalization group renormalization group invariance extends also to the case of quasi-polynomial velocity functions:

$$v_i(x_1,\ldots,x_n) = \sum_{j=1}^{q_i} v_{ij0} \prod_{k=1}^{n} x_k^{g_{ijk}} \quad , \quad i=1,\ldots,n$$

In fact, if we now apply the renormalization group invariance condition, namely $v_i^*(x_1^*,\ldots,x_n^*) = v_i(x_1,\ldots,x_n)$, after some calculations we find the following compatibility conditions:

$$\sum_{k=1}^{n} g_{ijk} = \frac{d}{3-d} \quad , \quad 0 < d < 3, \quad \forall\, i, j \tag{5}$$

This is obviously a generalization of (4). Thus, from (5) we see that every power-law making up every quasipolynomial velocity function $v_i$ must verify independently condition (4). Consequently, in particular (5) implies that all quasi-polynomial functions $v_i$ must be homogeneous of the same degree $d/(3-d)$. Therefore, when modeling the reactions of $n$ metabolites, the renormalization group invariance criterion is compatible with the use of the actual power-law models (GMA systems in general, including S-system models in particular).

For the sake of brevity, calculations involving alternative functional forms (such as $n$-variable rational functions generalizing Michaelis-Menten or Hill rate-laws) are not presented here. Such calculations are direct generalizations of those given for a single reacting metabolite. Moreover, the conclusions are entirely similar, as far as those functions are not compatible with renormalization group invariance. We thus conclude that power-law and quasipolynomial rate-laws appear as the natural renormalization group invariant solutions in the Zero-Order renormalization group approximation.

*2.3. Alternative interpretation of the Zero-Order renormalization group exact solutions*

Let us consider a general power-law based model of a metabolic pathway, namely a description in terms of a system of differential equations of GMA form, which for simplicity we can write as:

$$\frac{dx_i}{dt} = v_i(x_1,\ldots,x_n) = \sum_{j=1}^{q_i} v_{ij}(x_1,\ldots,x_n) = \sum_{j=1}^{q_i} v_{ij0} \prod_{k=1}^{n} x_k^{g_{ijk}} \quad , \quad i=1,\ldots,n \tag{6}$$

Therefore in (6) we have that $v_{ij}(x_1,\ldots,x_n)$ is a power-law for all *i, j*. In addition, we assume that the renormalization group condition (4) is verified by every power-law function $v_{ij}$ or equivalently, that identity (5) holds for all functions $v_i$ conforming the right-hand side of equations (6). Consider now that a scaling transformation is applied over the GMA system (6). Such transformation is defined in terms of a one-parameter Lie group transforming the system variables as: $y_i = \mu x_i$ for all $i=1,\ldots,n$. Since the $x_i$ are the system concentrations, actually such scaling is not a geometric one (as it was before, in the application of the renormalization group criterion) but a scaling in the metabolite concentrations. This explains that the group parameter now is termed $\mu$ instead of $\lambda$. In the new, transformed variables, the GMA system becomes:

$$\frac{dy_i}{dt} = \mu \frac{dx_i}{dt} = \mu \sum_{j=1}^{q_i} v_{ij}(x_1,\ldots,x_n) = \mu \sum_{j=1}^{q_i} v_{ij}(\mu^{-1}y_1,\ldots,\mu^{-1}y_n) = \mu^{1-\Gamma} \sum_{j=1}^{q_i} v_{ij}(y_1,\ldots,y_n)$$

Precisely, the last step in the previous equation is the direct consequence of identities (4) and (5): note that if (5) holds for a power-law function $v_{ij}$, then after some direct calculations we find that $v_{ij}(\mu^{-1}y_1,\ldots,\mu^{-1}y_n) = \mu^{-\Gamma} v_{ij}(y_1,\ldots,y_n)$. Therefore, in the new variables $y_i$ the GMA system remains the same with the only exception of a constant multiplicative factor $\mu^{1-\Gamma}$. This means that if the GMA system velocities $v_i$ in (6) verify the renormalization group condition (5), then the phase-space trajectories of the rescaled flow are identical to the original ones, namely such trajectories are exactly the same. In fact, this can be shown in a simple way if we perform a time reparametrization of the form $d\tau = \mu^{1-\Gamma} dt$, where *t* is the initial time variable and $\tau$ is the new time. We thus see that:

$$\frac{dy_i}{d\tau} = \frac{dt}{d\tau} \cdot \frac{dy_i}{dt} = \mu^{\Gamma-1} \frac{dy_i}{dt} = \sum_{j=1}^{q_i} v_{ij}(y_1,\ldots,y_n) \quad , \quad i=1,\ldots,n \tag{7}$$

As indicated, a time reparametrization does not modify the system trajectories, its only effect being a rescaling of the "speed" at which the phase state vector moves over such trajectories. Therefore, we see in (7) that the phase-space of the GMA system remains invariant after the scaling in the concentrations, and this happens if and only if the Zero-Order renormalization group invariance condition is verified. Mathematically, such condition in fact means that the GMA system is homogeneous, as anticipated, and this property accounts for the invariance found. Note that this property is independent of the actual value of parameter $\Gamma$, actually. Then, we have shown that the renormalization group invariance implies also phase-space invariance under scaling of concentrations and, in particular, this means that the GMA model remains valid over an infinite range of concentrations. In fact, this result accounts for the property, already mentioned in the Introduction, involving the reported validity of GMA and S-system models over several orders of magnitude in concentrations. In the Zero-Order approximation we obtain an idealized result in which the theoretical renormalization group range for concentrations is infinite, something that will be polished in the next section. We thus see how the renormalization group methods provide a direct link between self-similarity (geometric scaling) and the validity of the model over broad variations in the system variables (scaling in metabolite concentrations). Therefore the renormalization group approach leads to a unified perspective of both aspects of power-law modeling.

## 3. First-Order renormalization group approximation: exact solution

As indicated in the Introduction, the renormalization group is essentially an approximation, as are the corresponding results. The Zero-Order renormalization group calculations just developed provide a good starting point in order to understand the significant role of power-law models in the implementation and description of self-similarity. Actually, the results are valid for all values of λ, therefore describing a perfect self-similarity at all scales. This is clearly an idealization. It seems thus a good idea to proceed to a higher degree of approximation, in order to refine and validate different aspects of the results previously found. As it was done in the last section, now two cases (a single metabolite and more than one metabolite) are distinguished for the sake of clarity.

*3.1. One-variable case*

As in Subsection 2.1, we consider a single metabolite *A* reacting in a volume *V* and fractal dimension *d*. With the same notation, *M* is the initial number of molecules in that volume, and *N* is the increment per unit of time in the number of molecules of *A* in that region. As before, the velocity is *v=N/V*. After a geometric scaling of factor λ, we again have $V^* = \lambda^3 V$ as well as $M^* = \lambda^d M$. These are general features that remain unaltered. Now the refinement to be presented arises from the evaluation of the concentration *x*. In the Zero-Order renormalization group approach, it was *x=M/V*. However, *M* is the initial number of molecules in the region, namely such number ignores the variation due to the reaction taking place. In order to refine the evaluation of *x*, this time we shall consider the evolution due to the ongoing reaction during an infinitesimal time interval δ. At the initial time *t*, it is true that the concentration in the region is *M/V*. However at the time t+δ, such concentration will be *M/V + δv*. We can thus make use of an averaged value, and write in what follows that:

$$x = \frac{M}{V} + \frac{N}{V}\frac{\delta}{2} = \frac{M}{V}\left(1 + \frac{N}{M}\frac{\delta}{2}\right) \equiv \frac{M}{V}(1+\varepsilon) \qquad (8)$$

where ε<<1 is a small dimensionless parameter. The smallness of ε arises from the fact that typically N<<M and, in addition, the time interval δ is infinitesimal and for practical purposes can be made arbitrarily small. It is worth recalling that the kind of approximation based on taking an average value at the central point t+δ/2 of the time interval is very common in numerical analysis, for instance in the domain of finite element methods (Young and Gregory, 1973). Notice also from (8) that in the limit case $\varepsilon \to 0$, we remain in the Zero-Order renormalization group approximation. As before, we now assume that the geometric and dynamic invariance of the system in the new scale can be quantitatively expressed as:

$$\frac{N}{MV} = \frac{N^*}{M^*V^*}$$

Taking into account this invariance condition, we again obtain $N^* = \lambda^{d+3} N$ and $v^* = N^*/V^* = \lambda^d v$. Finally, we can use (8) in order to compute $x^*$:

$$x^* = \frac{M^*}{V^*}\left(1 + \frac{N^*}{M^*}\frac{\delta}{2}\right) = \lambda^{d-3}\frac{M}{V}\left(1 + \lambda^3 \varepsilon\right)$$

Let us now impose the one-variable renormalization group condition for velocities, namely $v^*(x^*) = v(x)$. In what follows, we shall focus on the case of a power-law rate-law, $v(x) = v_0 x^g$. Then, after some simplifications we arrive to the condition: $(1+\varepsilon)^g = \lambda^{d+g(d-3)}\left(1 + \lambda^3 \varepsilon\right)^g$. Note that this equation is identically satisfied when λ=1 for all values of g, d and ε. In addition, in the limit $\varepsilon \to 0$, we consistently retrieve the Zero-Order renormalization group identity (2), as expected. From the previous First-Order renormalization group relationship, by taking logarithms we obtain that:

$$g = \frac{d}{(3-d) + \frac{1}{\log \lambda}\log\left(\frac{1+\varepsilon}{1+\lambda^3 \varepsilon}\right)} \quad (9)$$

As before, the limit $\varepsilon \to 0$ amounts to the Zero-Order renormalization group identity (2). However, now some important differences arise. First of all, notice that now the Zero-Order divergence observed for $d=3$ is avoided. In fact, $g$ can take a wide range of values for $d \leq 3$ without a formal divergence. As an illustration of this result, Figure 1 displays a typical behavior of the First-Order renormalization group kinetic order (9) for a fixed value of $\lambda$ and varying values of $d$ and $\varepsilon$.

**Figure 1.**

An additional point of view regarding equation (9) is provided by Figure 2. In this case, the First-Order renormalization group kinetic order (9) is displayed for a fixed value of $d$ and variable values of $\lambda$ and $\varepsilon$.

**Figure 2.**

Figure 2 also illustrates another relevant property of the First-Order renormalization group result (9), namely that such kinetic order is dependent on $\lambda$. As explained before, an exact renormalization group invariant solution being independent of $\lambda$, as it was the case in the Zero-Order approach, reflects a perfect and exact invariance at all scales. However now this is not the case, as far as (9) is $\lambda$-dependent. This means that, in fact, a given power-law rate-law now can be an exact solution of the First-Order renormalization group equations for a single value $\lambda_0$, but not for all values of $\lambda$. In other words, in this approximation a power-law with a fixed kinetic order $g$ does not comply to a perfect self-similarity at all scales. However, this is not the end of the story. Actually, power-law functions do present an approximate First-Order renormalization group invariance for a finite range of scales: the reason is that condition (9) depends on the small parameter $\varepsilon$. As indicated, if $\varepsilon=0$ we have an exact invariance for all scales (Zero-Order approximation). Thus, in the First-Order approach the power-law is not exactly renormalization group invariant for every $\lambda$, but it is invariant to a high degree of approximation over an interval of values of $\lambda$, which depends on the small parameter $\varepsilon$. This is the case because function $\log\left(\frac{1+\varepsilon}{1+\lambda^3\varepsilon}\right)$ in (9) is close to zero in absolute value and varies slowly as $\lambda$ changes. For instance, Figure 3 presents the relative deviation between the First-Order kinetic order (9) and the Zero-Order one (2). Let us denote both kinetic orders by $g_F$ and $g_Z$, respectively. Then the relative deviation is defined as $R=(g_Z-g_F)/g_Z$. Accordingly, Figure 3 depicts the variation of $R$ for a fixed value of the fractal dimension $d$ and for broad variations in the scaling parameter $\lambda$ and in $\varepsilon$. As expected, the relative variation remains small.

**Figure 3.**

Therefore, it is worth recalling that the degree of approximate First-Order renormalization group invariance of a power-law is dependent on the precise value of $\lambda$, not being equal at all scales. As we have seen, now $g$ is a function of $(d, \lambda, \varepsilon)$. This means that the renormalization group invariant exponent $g$ now is not one and the same for all scales. Accordingly, a given power-law can only provide exact invariance for a given scaling, namely for a given value $\lambda_0$ of parameter $\lambda$, as well as a very approximate invariance for values of $\lambda$ close to $\lambda_0$ (for constant values of $d$ and $\varepsilon$). Emphasizing once again the approximate character of the renormalization group techniques, the previous conclusions seem to provide a justification about the prominent role of the power-law functions in the description of metabolic pathways in self-similar media: briefly speaking, it can be concluded that such functions approximate very precisely the property of self-similarity over a broad (but necessarily finite) range of scales.

In what follows, let us analyze the generalization to several reacting metabolites.

*3.2. Generalization to the n-variable case*

As it was the case in Subsection 2.2, we again consider the more general situation in which several metabolites $A_1,...,A_n$ react in a region of volume $V$ and fractal dimension $d$. As before, $M_i$ is the initial number of molecules of $A_i$ in that volume for every $i=1,...,n$, respectively. Also, $N_i$ is the increment in the number of molecules of $A_i$ in that region per unit of time, and then $v_i=N_i/V$. Accordingly, if the system remains self-similar after a scaling of factor $\lambda$, we again have $V^*= \lambda^3 V$ and $M_i^*= \lambda^d M_i$. As in the previous arguments, it is postulated that the dynamical invariance of the system in the new scale can be quantitatively expressed as:

$$\frac{N_i}{M_i V} = \frac{N_i^*}{M_i^* V^*}$$

for all $i=1,...,n$. After such relationship we again find $N_i^*= \lambda^{d+3} N_i$, as well as $v_i^*= \lambda^d v_i$. Recall that this condition must be written in functional terms as $v_i^*(\xi_1,...,\xi_n)= \lambda^d v_i(\xi_1,...,\xi_n)$ for any $n$-variable argument $(\xi_1,...,\xi_n)$. To complete the preliminaries, we need an expression for the concentrations $x_i$. Following a similar reasoning to that in Subsection 3.1, we introduce an infinitesimal time interval $\delta$ and write an averaged value for the concentrations:

$$x_i = \frac{M_i}{V} + \frac{N_i}{V}\frac{\delta}{2} = \frac{M_i}{V}\left(1+\frac{N_i}{M_i}\frac{\delta}{2}\right) \equiv \frac{M_i}{V}(1+\varepsilon_i)$$

Consequently, after the scaling the concentrations are transformed as:

$$x_i^* = \frac{M_i^*}{V^*}\left(1 + \frac{N_i^*}{M_i^*}\frac{\delta}{2}\right) = \lambda^{d-3}\frac{M_i}{V}\left(1 + \lambda^3 \varepsilon_i\right)$$

Once again, we wish to investigate the renormalization group invariance. As we know, the renormalization group criterion amounts to $v_i^*(x_1^*,\ldots,x_n^*) = v_i(x_1,\ldots,x_n)$ for all $i=1,\ldots,n$. Specifically, we shall analyze the case of a power-law functional dependence, i.e. we investigate the invariance of the following set of rate-laws:

$$v_i(x_1,\ldots,x_n) = v_{i0}\prod_{j=1}^{n} x_j^{g_{ij}} \quad , \quad i=1,\ldots,n$$

After some calculations based on the same procedures used in previous sections, eventually we arrive to the following set of compatibility conditions:

$$\sum_{j=1}^{n}\left[3 - d + \frac{1}{\log\lambda}\log\left(\frac{1+\varepsilon_j}{1+\lambda^3\varepsilon_j}\right)\right]g_{ij} = d, \quad i=1,\ldots,n \tag{10}$$

Notice that these equations admit an infinity of possible solutions for the kinetic orders $g_{ij}$, including in particular the possibility of solutions having negative components. In addition, in the limit $\varepsilon_i \to 0$ the Zero-Order renormalization group result (4) is retrieved. The nature of relationships (10) is more evident if we consider the particular case in which $\varepsilon_1 = \ldots = \varepsilon_n \equiv \varepsilon$. If this is the situation, equation (10) becomes simplified as:

$$\sum_{j=1}^{n}g_{ij} \equiv \Gamma = \frac{d}{3 - d + \dfrac{1}{\log\lambda}\log\left(\dfrac{1+\varepsilon}{1+\lambda^3\varepsilon}\right)}, \quad i=1,\ldots,n \tag{11}$$

It is evident in (11) that it is a direct generalization of (9). It is thus natural that the conclusions obtained in the one-variable First-Order renormalization group approach remain valid.

As before, the limit $\varepsilon \to 0$ amounts to the Zero-Order renormalization group identity (4). However, now the Zero-Order divergence observed for $d=3$ is again avoided, and the kinetic orders can take a broad range of values also in the case $d=3$. The second relevant property of the First-Order renormalization group result (11) is that such equations are again dependent on $\lambda$, thus showing that power-law rate-laws can be exact solutions of the First-Order renormalization group equations for a given value $\lambda_0$ of the scaling parameter $\lambda$, but not for all values of $\lambda$. In addition, we now have an approximate renormalization group invariance for values of $\lambda$ close to $\lambda_0$, and such approximate invariance is due to the dependence on the small parameter $\varepsilon$. Then, as it was the case in the one variable First-Order renormalization group analysis, the power-law functional form is not exactly renormalization group invariant over a finite interval of $\lambda$ values, but it is invariant to a high degree of approximation over such interval which depends on the small parameter $\varepsilon$. In addition, the dependence on $\lambda$ implies that the First-Order renormalization group exponents $g_{ij}$ now are not the same for all scales. Accordingly, a given set of power-laws can only provide an exact invariance for a certain scaling parameter $\lambda_0$ as well as an approximate invariance for values of $\lambda$ close to $\lambda_0$ (provided constant values of $d$ and $\varepsilon$). Therefore, we again conclude that power-law rate-laws implement very precisely the property of self-similarity over a limited (not infinite) range of scales. As shown, in the limit $\varepsilon \to 0$ the exact (infinite) invariance is retrieved. Therefore, the property of self-similarity over a limited range of scales is implemented with an accuracy which can be as precise as desired if $\varepsilon$ takes a value sufficiently close to zero in absolute value. Obviously, following a reasoning similar to the one displayed in Subsection 2.2, analogous conclusions are valid for quasi-polynomial rate-laws. However, the calculations are omitted for the sake of conciseness.

*3.3. Alternative interpretation of the First-Order renormalization group exact solutions*

As it was shown in the Zero-Order renormalization group analysis, when applied to a general GMA model (6), conditions (4) and (5) amount to the fact that the GMA differential equations are homogeneous. This allowed an additional interpretation of the Zero-Order renormalization group condition as the GMA system invariance with respect to the scaling group $y_i=\mu x_i$, $i=1,...,n$. As indicated, since the $x_i$ are the system variables (which are not lengths), actually such scaling is not a geometric one (as it was before, in the application of the renormalization group criterion) but a scaling in the metabolite concentrations. In the First-Order renormalization group approximation, now conditions (4) are replaced by their generalization (10). As we have seen, such generalization depends on the small parameters $\varepsilon_i$ in such a way that (4) is retrieved in the limit $\varepsilon_i \to 0$. Due to the smallness of parameters $\varepsilon_i$ it is clear that conditions (10) imply that the associated GMA models are almost homogeneous, the exact homogeneity arising only in the limit $\varepsilon_i \to 0$. Accordingly, the First-Order renormalization group criterion leads to an approximate GMA system invariance with respect to the scaling group $y_i=\mu x_i$, $i=1,...,n$. In other words, we have shown that GMA models verifying (10), namely almost homogeneous GMA systems, remain valid over a broad (but this time finite) range of concentrations. According to the original renormalization group hypothesis this result provides a link for GMA systems between (i) an approximate geometric self-similarity of the system over a finite range of geometric scaling factors, and (ii) an approximate phase-space invariance under scaling of concentrations over a finite range of variation of such concentrations. Consequently, we retrieve the property, mentioned in the Introduction, involving the validity of GMA and S-system models over several orders of magnitude in concentrations. In the Zero-Order approximation we obtained an idealized result in which the theoretical renormalization group range for concentrations is infinite, something that now has been polished. Therefore the First-Order renormalization group approximation allows establishing a direct link between finite-range self-similarity (geometric scaling) and the validity of the model over broad variations of the system variables (scaling in metabolite concentrations).

*3.4. Example: linear chain with feedback inhibition*

In what follows a detailed example is presented for the sake of illustration. Specifically, the metabolic pathway considered is a linear chain with feedback control by inhibition, a well-known mechanism (Savageau, 1976). Such pathway is schematically displayed in Figure 4.

---

**Figure 4.**

---

In terms of power-law modeling, and in agreement with equations (4) and (5), the Zero-Order model for the system is described by the following set of equations, in which for every index $i$ the variable $x_i$ accounts for the concentration of the metabolite $X_i$.

$$\frac{dx_1}{dt} = a_1 x_0^{\Gamma+\gamma} x_2^{-\gamma} - b_1 x_1^{\Gamma}$$
$$\frac{dx_2}{dt} = a_2 x_1^{\Gamma} - b_2 x_2^{\Gamma}$$
(12)

Recall that this Zero-Order model is necessarily homogeneous, as discussed in Section 2. We see that the resulting GMA model (12) is also an S-system. Notice that, as usual in this case, the concentration $x_0$ of the initial precursor $X_0$ is considered to have a fixed value that can be independently modified by experimental means. The resulting system is thus two-dimensional, a very convenient feature for illustrative purposes, as we shall see. In what follows, the values $\Gamma = 2.5$, $d = 15/7 \approx 2.14$ and $\gamma = 0.5$ will be the ones employed. In addition, for the sake of clarity it will be very convenient to have the steady-state of the Zero-Order model in the coordinates (1,1). Accordingly, we set $a_1 x_0^{\Gamma+\gamma} = b_1 = a_2 = b_2 = 1$. This is just a convenient choice of parametric values suitable for presentation purposes, but any other set of values would be equally good. The dynamics of the resulting system correspond to a stable steady-state, as depicted in Figure 5.

**Figure 5.**

Let us now define the First-Order system associated to this pathway. For this, we must take into account equation (10). For convenience, let us introduce the following definition:

$$w_j(d, \lambda, \varepsilon_j) = 3 - d + \frac{1}{\log \lambda} \log\left(\frac{1+\varepsilon_j}{1+\lambda^3 \varepsilon_j}\right), \quad j = 0, 1, 2$$

We then arrive to the First-Order system, which is given by:

$$\frac{dx_1}{dt} = a_1 x_0^{\Gamma+\gamma} x_2^{(d-(\Gamma+\gamma)w_0)/w_2} - x_1^{d/w_1}$$
$$\frac{dx_2}{dt} = x_1^{d/w_1} - x_2^{d/w_2}$$
(13)

As it can be seen, in the previous equations the kinetic order of $X_0$ is maintained as $\Gamma+\gamma$ for clarity of illustration as in the Zero-Order case (recall that by definition it is $a_1 x_0^{\Gamma+\gamma} = 1$). Note that the expression $a_1 x_0^{\Gamma+\gamma}$ takes the value 1 for convenience, in order to place the steady-state at point (1,1), as mentioned before. Such factor is explicitly written in Eq. (13) for clarity, because from condition (10) the value of the exponent of $x_0$ determines the value of the exponent of $x_2$ in the same monomial. Alternatively, the factor $a_1 x_0^{\Gamma+\gamma}$ can be replaced by its constant value 1. Note that in general, the rescaling is applied to both independent and dependent variables: however, independent variables are actually constant parameters and, in fact, they remain unaltered when the rescaling is applied to them. This can be verified easily, and $x_0$ in the present model is an example of it. In addition we shall set $\varepsilon_0 = 0$ and consistently $w_0 = 3-d$. Also, the values $\varepsilon_1 = 10^{-5}$ and $\varepsilon_2 = -10^{-7}$ are defined for the rest of the example.

In first place, we shall consider the effect of geometric scalings over the First-Order system (13). According to (10), mathematically this amounts to the variation of parameter λ. Of course, this variation modifies the values of the kinetic orders in model (13). The effect of progressively higher scalings is displayed Figure 6. In particular, the phase plots are displayed for λ=2 (Figure 6a), λ= 50 (Figure 6b) and λ=100 (Figure 6c). In addition, the trajectories displayed in each plot are equivalent, namely for the sake of comparison the same initial conditions have been chosen for the trajectories plotted in Figure 5 and in Figures 6a, 6b and 6c. The presentation of this kind of graphical comparison is possible as far as the steady-state remains by construction located at (1,1), as anticipated. It is evident that even with a 100-fold geometric scaling (namely, over several orders of magnitude) the system trajectories remain essentially unaltered, in full agreement with the approximate invariance established for the First-Order renormalization group power-law models.

**Figure 6 (Figures 6a, 6b, 6c).**

In order to complete the example, we shall now turn to examine the effect of scalings in the metabolite concentrations in the First-Order system (13). Let us recall that such scalings are given by the one-parameter Lie group $y_i = \mu x_i$ for all $i=1,2,3$. When applied to system (13), and taking into account that $a_1 x_0^{\Gamma+\gamma} = 1$, the outcome is:

$$\frac{dy_1}{dt} = \mu^{1-(d-(\Gamma+\gamma)w_0)/w_2} y_2^{(d-(\Gamma+\gamma)w_0)/w_2} - \mu^{1-d/w_1} y_1^{d/w_1}$$

$$\frac{dy_2}{dt} = \mu^{1-d/w_1} y_1^{d/w_1} - \mu^{1-d/w_2} y_2^{d/w_2}$$
(14)

By means of system (14) we can check the additional effect of scaling in concentrations, which is displayed in Figure 7.

**Figure 7 (Figures 7a, 7b, 7c).**

More precisely, the phase plots in Figure 7 are displayed for the values $\mu=2$ (Figure 7a), $\mu=50$ (Figure 7b) and $\mu=100$ (Figure 7c). In all cases, the geometric scaling parameter has the value $\lambda=2$. Note that, consistently, for every value of parameter $\mu$ now the steady-state is approximately centered at the coordinates $(\mu,\mu)$. Again, even with a 100-fold concentration scaling it is clear that near-invariance is present as far as the stable steady-state topology of the solutions remains unaltered. The "straight-line" aspect of trajectories in Figures 7b and 7c is due to the magnification efect produced on the phase-plot by this kind of transformation. Consequently, we verify also in the case of concentration scalings the approximate invariance established for the First-Order renormalization group power-law models over several orders of magnitude.

## 4. Discussion

As shown throughout the previous section, power-law functions can be regarded as the most basic renormalization group invariant rate-laws. This has several implications that deserve some additional comments.

In first place, it is worth recalling that geometric invariance is actually present only over a limited (not infinite) range of scales. Precisely, the significance of power-laws is that they approximate such invariance over a limited range of scales, which is the actual property to be described. As we have seen, such feature is not present in other common rate-laws such as Michaelis-Menten, Hill, or their generalizations. However, the previous results are also useful in order to account for the ability of many functions to provide accurate predictions when employed in the construction of differential models: a possible explanation is that such functions are often relatively close to power-laws (Sorribas *et al.* 2007); accordingly, those functions could, in some situations, reproduce precisely the actual invariance properties of the system which is being analyzed.

It is also convenient to provide some additional considerations regarding the assumptions being made in the present work. This is necessary in order to put in perspective the present results with respect to previous approaches (Kopelman, 1986, 1991; Savageau, 1993, 1995, 1998). In first place, it is worth noting that the existence of kinetic equilibrium is not assumed, as far as an ongoing reaction is considered in the previous calculations. As a consequence, the concentrations are in fact not taken as constant in the first-order analysis. The fact that the average kinetic order is a decreasing function of the fractal dimension $d$, as well as the precise functional form of this dependence, is one the most direct and novel consequences obtained from this point of view. Another difference with previous approaches is that usually the effects of a dimensionally limited diffusion are the ones analyzed, while here we have focused on purely reactional effects (namely excluding diffusion). In other words, our approach deals with kinetics taking place in a dimensionally limited environment, that is to say a region of fractal dimension $d < 3$, on which only reaction kinetics is considered. In the present work, the specific analysis of reactional effects has allowed the use of the renormalization group methods in order to develop the combined perspective of geometric scalings and concentration scalings just presented, which is to the author's knowledge new in the literature. In addition, it seems that focusing on reactional effects is natural in this context, as far as power-law models such as S-systems and GMA systems are purely reactional.

Another aspect of interest involves the practical use of GMA models and the evaluation of kinetic orders. As we have seen in the First-Order renormalization group approach, according to (10) the invariant GMA models should be almost homogeneous. In practice, it is worth recalling the difficulties in the construction of models (e.g. see Voit, 2000). In fact, even the fitting of a single power-law from data is today a controversial issue (Kaitaniemi, 2004; Packard, 2009). Due to the relevance of different uncertainties and sources of errors (experimental and numerical) in parameter estimation for model determination, specially when modeling *in vivo* processes, probably it is not to be expected that the homogeneity or almost homogeneity properties should be apparent in practice.

As recalled in the previous analysis, the renormalization group methodology is an approximation. In spite that its quantitative predictions are often not precise, renormalization group equations usually provide qualitative hints of great value in a workable way. This explains the successes of renormalization group theory in many different domains. In our case, the most important results obtained are those providing a link among three aspects of modeling: (i) the use of power-laws for the description of rate-laws; (ii) the geometric (and according to relationships (3) also physical) self-similarity under spatial scalings; and (iii) the self-similarity under concentration scalings (accounting for the reported GMA model validity over several orders of magnitude in metabolite concentrations). In particular, the last item is a consequence of the system invariance characterized which provides a mathematical background for the observed model validity over different orders of magnitude in concentrations. This feature was presented in detail in the Introduction, mathematically developed in Subsections 2.3 and 3.3, and illustrated in the example of Subsection 3.4. The conceptual link thus established between power-law modeling, fractality and different types of invariance (spatial scalings and concentration scalings) probably will lead to novel perspectives and tools from the modeling point of view, in spite that at the present stage the results obtained are essentially theoretical. In this sense, and on the basis of the assumptions established in order to develop the previous calculations, the qualitative results obtained (already discussed in detail in Sections 2 and 3) seem to provide additional insight on some fundamental reasons accounting for the successes of power-law modeling.

**Acknowledgements**

This work has been supported by the Spanish MEC, Project Ref. MTM2006-10053. In addition, the author acknowledges two anonymous referees for their advice and fruitful comments.

# References

Aon, M.A., O'Rourke, B., Cortassa, S., 2004. The fractal architecture of cytoplasmatic organization: Scaling, kinetics and emergence in metabolic networks. Mol. Cell. Biochem. 256/257, 169-184.

Auffray, C., Nottale, L., 2008. Scale relativity theory and integrative systems biology: 1 Founding principles and scale laws. Prog. Biophys. Molec. Biol. 97, 79-114.

Creswick, R.J., Farach, H.A., Poole, C.P., 1992. Introduction to renormalization group methods in physics. New York: John Wiley & Sons.

Demetrius, L., 2006. The origin of allometric scaling laws in biology. J. Theor. Biol. 243, 455-467.

Derome, J.R., 1977. Biological similarity and group theory. J. Theor. Biol. 65, 369-378.

Díaz-Sierra, R., Hernández-Bermejo, B., Fairén, V., 1999. Graph-theoretic description of the interplay between non-linearity and connectivity in biological systems. Math. Biosci. 156, 229-253.

Fisher, M.E., 1998. Renormalization group theory: its basis and formulation in statistical physics. Rev. Mod. Phys. 70, 653-681.

Furusawa, C., Kaneko, K., 2006. Evolutionary origin of power-laws in a biochemical reaction network: embedding the distribution of abundance into topology. Phys. Rev. E 73, 011912.

Hernández-Bermejo, B., Fairén, V., 1997. Lotka-Volterra representation of general nonlinear systems. Math. Biosci. 140, 1-32.

Huxley, J.S., 1932. Problems of relative growth. New York: The Dial Press.

Kaitaniemi, P., 2004. Testing the allometric scaling laws. J. Theor. Biol. 228, 149-153.

Kopelman, R., 1986. Rate processes on fractals: theory, simulations and experiments. J. Stat. Phys. 42, 185-200.

Kopelman, R., 1991. Reaction kinetics in restricted spaces. Isr. J. Chem. 31, 147-157.

Packard, G.C., 2009. On the use of logarithmic transformations in allometric analysis. J. Theor. Biol. 257, 515-518.

Savageau, M.A., 1976. Biochemical systems analysis: a study of function and design in molecular biology. Reading (Mass.): Addison-Wesley.

Savageau, M.A., 1979a. Growth of complex systems can be related to the properties of their underlying determinants. Proc. Natl. Acad. Sci. USA 76, 5413-5417.

Savageau, M.A., 1979b. Allometric morphogenesis of complex systems: derivation of the basic equations from first principles. Proc. Natl. Acad. Sci. USA 76, 6023-6025.

Savageau, M.A., 1993. Influence of fractal kinetics on molecular recognition. J. Molec. Recogn. 6, 149-157.

Savageau, M.A., 1995. Michaelis-Menten mechanism reconsidered: implications of fractal kynetics. J. Theor. Biol. 176, 115-124.

Savageau, M.A., 1998. Development of fractal kinetic theory for enzyme-catalysed reactions and implications for the design of biochemical pathways. BioSystems 47, 9-36.

Schroeder, M., 1991. Fractals, chaos, power-laws. New York: W. H. Freeman and Company.

Sorribas, A., Hernández-Bermejo, B., Vilaprinyó, E., Alves, R., 2007. Cooperativity and saturation in biochemical networks: a saturable formalism using Taylor series approximations. Biotechnol. Bioeng. 97, 1259-1277.

Sorribas, A., Savageau, M.A., 1989a. A comparison of variant theories of intact biochemical systems, 1. Enzyme-enzyme interactions and biochemical systems theory. Math. Biosci. 94, 161-193.

Sorribas, A., Savageau, M.A., 1989b. A comparison of variant theories of intact biochemical systems, 2. Flux oriented and metabolic control theories. Math. Biosci. 94, 195-238.

Sorribas, A., Savageau, M.A., 1989c. Strategies for representing metabolic pathways within biochemical systems theory: reversible pathways. Math. Biosci. 94, 239-269.

Stinchcombe, R.B., 1988. Phase transitions. In: Lundqvist, S., March, N.H., Tosi, M.P. (Eds.), Order and chaos in nonlinear physical systems. Plenum Press, New York, pp. 295-340.

Takayasu, H., 1990. Fractals in the physical sciences. Manchester (UK): Manchester University Press.

Voit, E.O., 1992. Symmetries of S-systems. Math. Biosci. 109, 19-37.

Voit, E.O., 2000. Computational analysis of biochemical systems: a practical guide for biochemists and molecular biologists. Cambridge (UK): Cambridge University Press.

Voit, E.O., Savageau, M.A., 1987. Accuracy of alternative representations for integrated biochemical systems. Biochemistry 26, 6869-6880.

Voit, E.O., Savageau, M.A., Irvine, D.H., 1991. Introduction to S-systems. In: Voit, E.O. (Ed.), Canonical nonlinear modeling. Van Nostrand Reinhold, New York, pp. 47-66.

West, B.J., 2004. Comments on the renormalization group, scaling and measures of complexity. Chaos, Solitons and Fractals 20, 33-44.

West, G.B., 1999. The origin of universal scaling laws in biology. Physica A 263, 104-113.

West, G.B., Woodruff, W.H., Brown, J.H., 2002. Allometric scaling of metabolic rate from molecules and mitochondria to cells and mammals. Proc. Natl. Acad. Sci. USA 99, 2473-2478.

West, G.B., Brown, J.H., 2005. The origin of allometric scaling laws in biology from genomes to ecosystems: towards a quantitative unifying theory of biological structure and organization. J. Experim. Biol. 208, 1575-1592.

Wu, R., Ma, C.-X., Littell, R.C., Casella, G., 2002. A statistical model for the genetic origin of allometric scaling laws in biology. J. Theor. Biol. 219, 121-135.

Young, D.M., Gregory, R.T., 1973. A survey of numerical mathematics, Vol. II. Mineola (New York): Dover Publications.

**FIGURE CAPTIONS**

**Figure 1:**

***First-order renormalization group kinetic orders.*** A typical behavior of the first-order renormalization group kinetic order (9) for some selected values. In this graph, the value $\lambda=5$ is the one selected for the scaling parameter. It can be seen that the kinetic order $g$ is a decreasing function of the fractal dimension $d$.

**Figure 2:**

***First-order renormalization group kinetic orders.*** Alternative representation of a typical behavior of the first-order renormalization group kinetic order (9) for some selected values. In this graph, the value $d=2.3$ of the fractal dimension is fixed, while $\lambda$ and $\varepsilon$ are allowed to vary.

**Figure 3:**

***Relative deviation between the renormalization group kinetic orders.*** Behavior of the relative deviation $R$ for the kinetic order $g$ between the first and Zero-Order renormalization group approximations. $R$ is displayed for a fixed value of the fractal dimension $d=2.3$ and for broad variations in the scaling parameter $\lambda$ and in $\varepsilon$. As expected, the relative deviation remains small, in this case within limits of $\pm 5\%$.

**Figure 4:**

***Metabolic pathway for the example in Section 3.4.*** The example considers a linear chain with feedback control by inhibition. The end product $X_2$ acts acts as an allosteric effector causing inhibition of the first reaction in the sequence.

**Figure 5:**

***Trajectories of the Zero-Order system (12).*** The phase plot shows the steady-state of system (12) which is placed by construction at point (1,1) and is stable. The parameter values of the system are detailed in the main text.

**Figure 6:**

***Trajectories of the First-Order system (13) for different values of the scaling parameter λ.*** The phase plots are displayed for λ=2 (Figure 6a), λ= 50 (Figure 6b) and λ=100 (Figure 6c). The trajectories displayed in each plot are equivalent, namely for the sake of comparison the same initial conditions have been chosen for the trajectories plotted in Figure 5 and in Figures 6a, 6b and 6c.

**Figure 7:**

***Trajectories of the First-Order system (14) for different values of the concentration scaling parameter μ.*** The phase plots are displayed for μ=2 (Figure 7a), μ= 50 (Figure 7b) and μ=100 (Figure 7c). In all cases, the geometric scaling parameter has the value λ=2.

**FIGURE 1:**

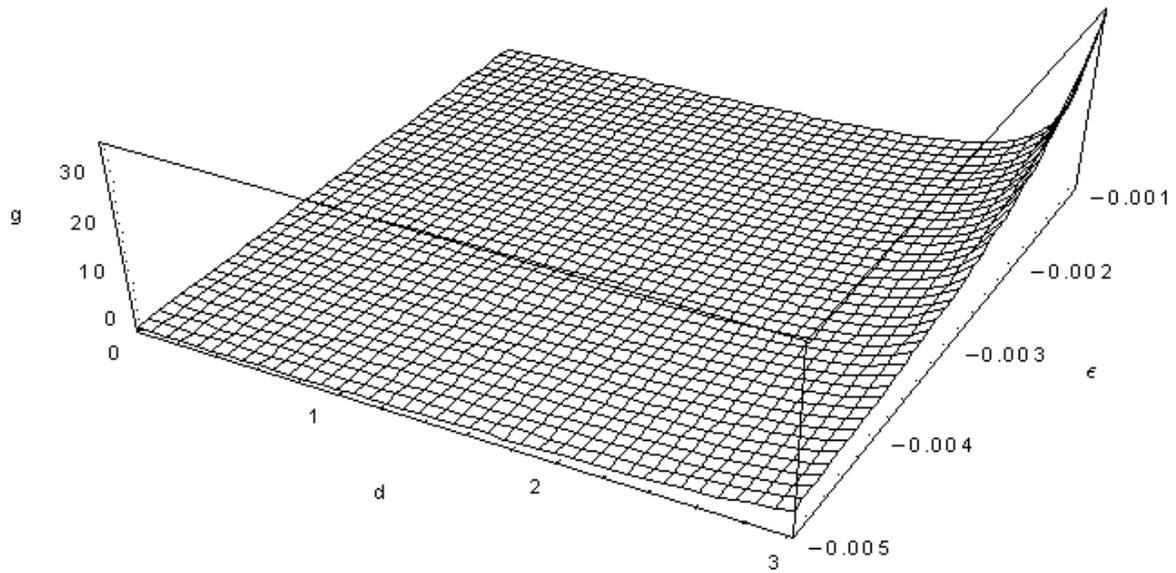

**FIGURE 2:**

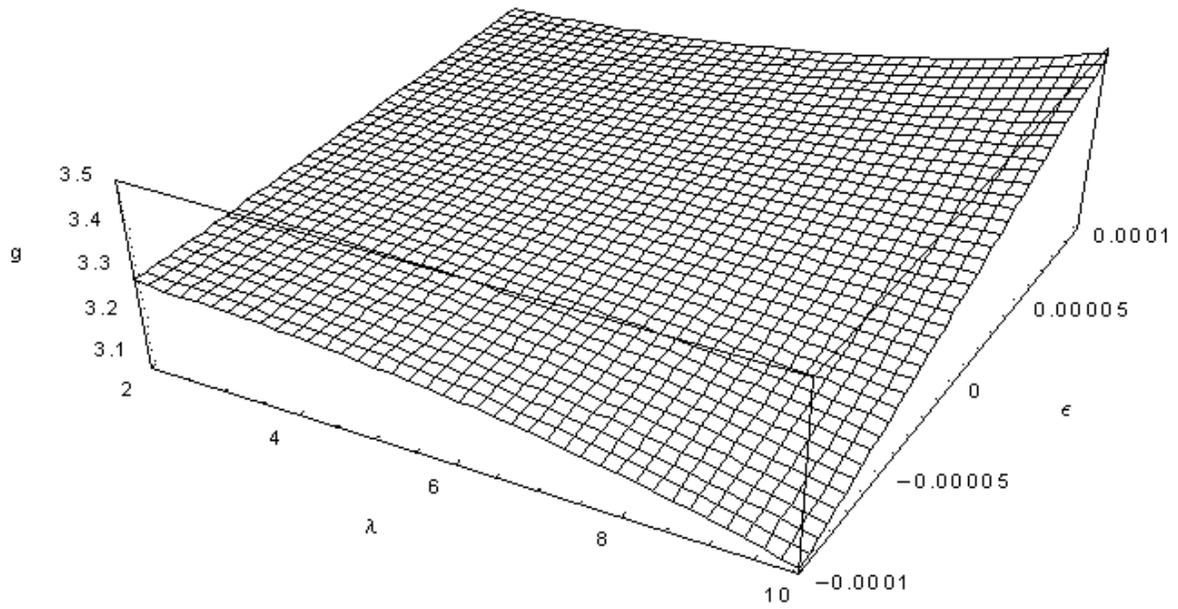

**FIGURE 3:**

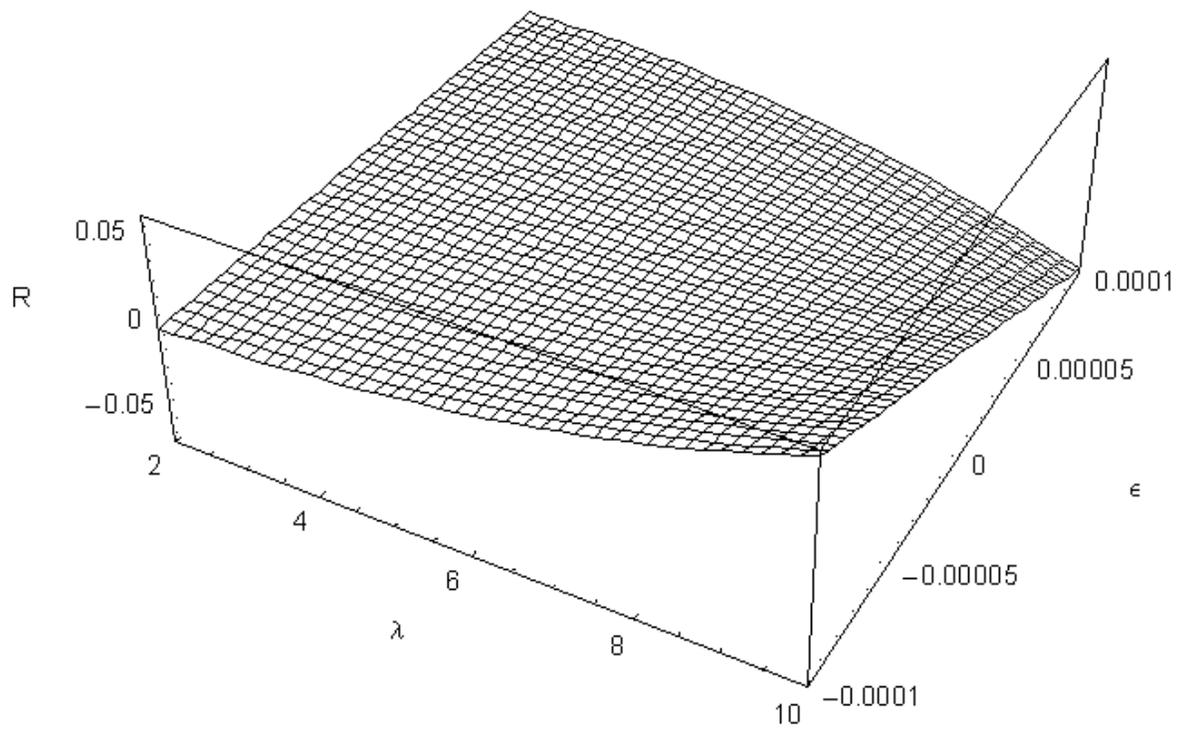

**FIGURE 4:**

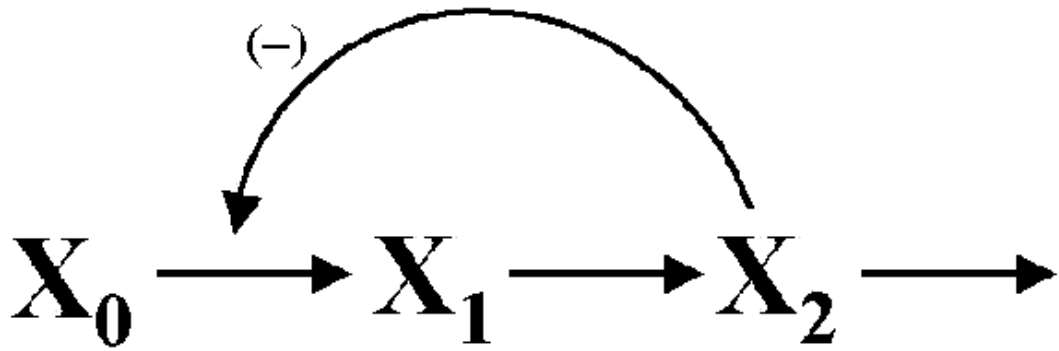

**FIGURE 5:**

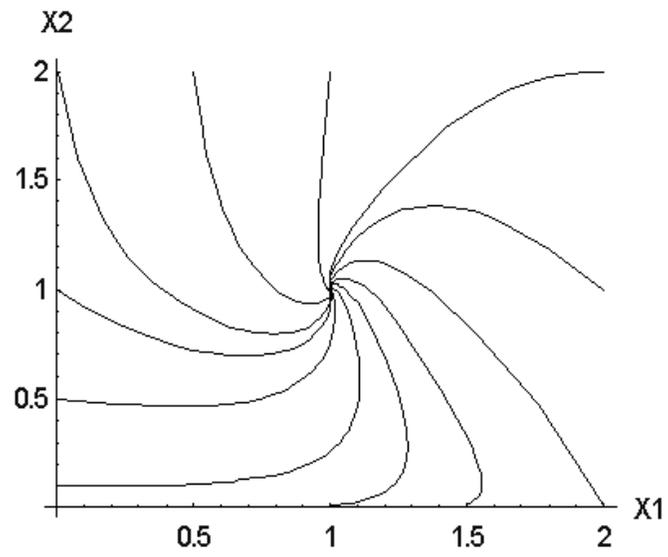

**FIGURE 6a:**

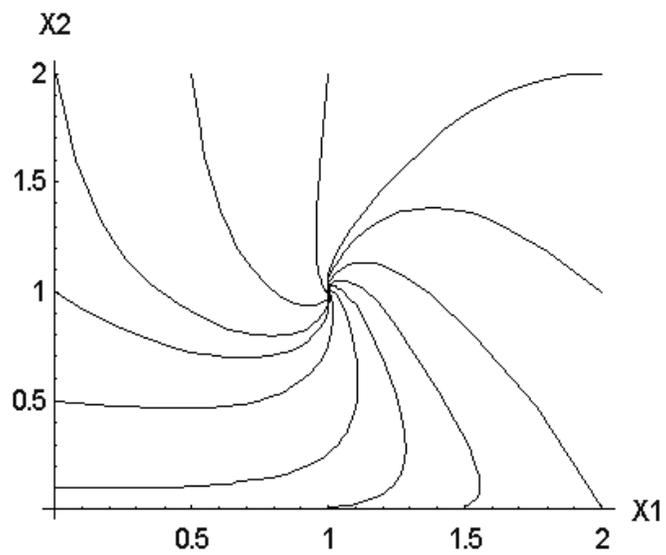

**FIGURE 6b:**

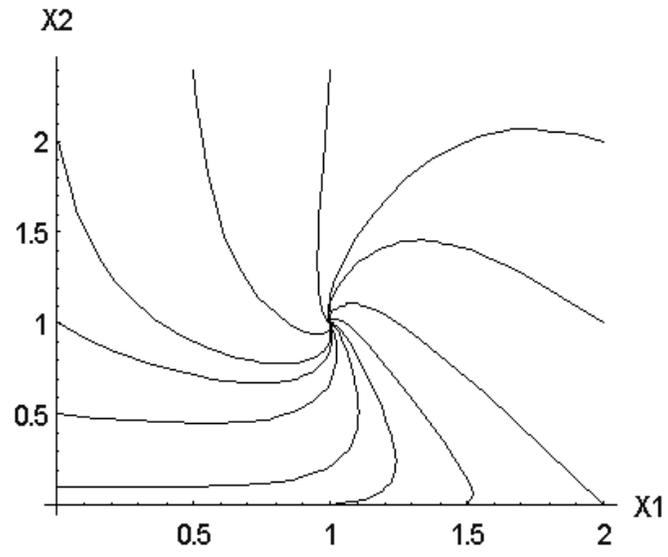

**FIGURE 6c:**

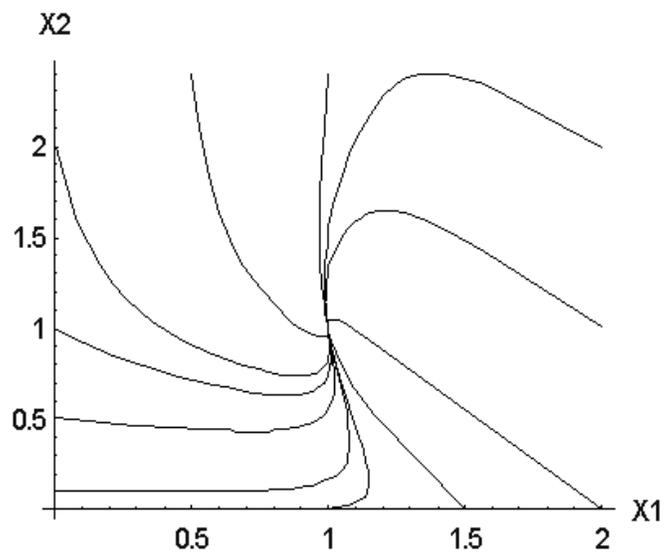

**FIGURE 7a:**

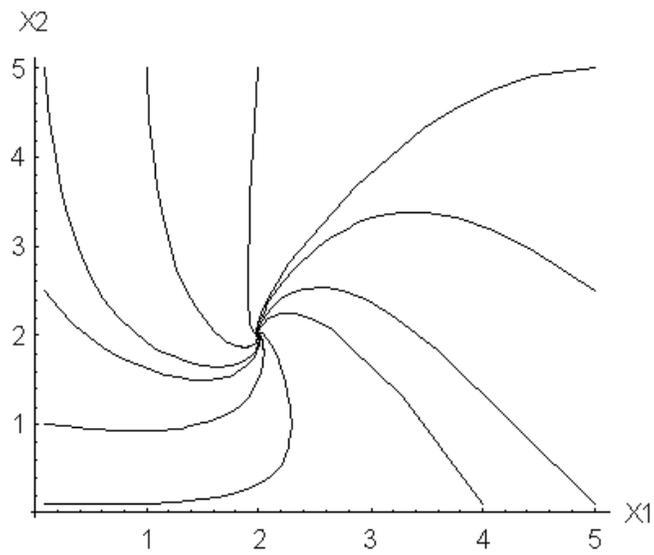

**FIGURE 7b:**

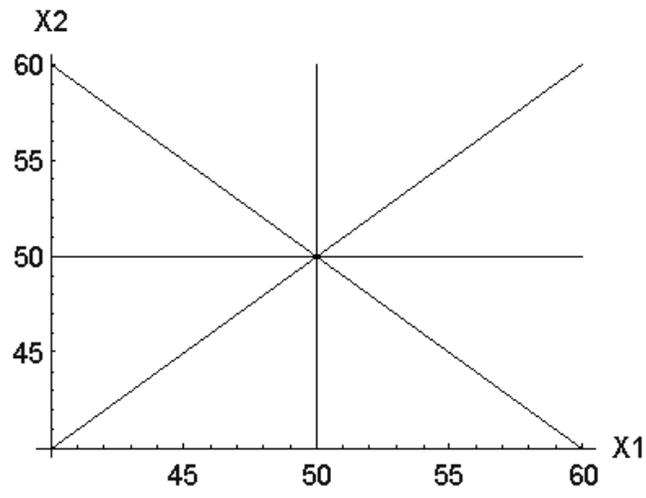

**FIGURE 7c:**

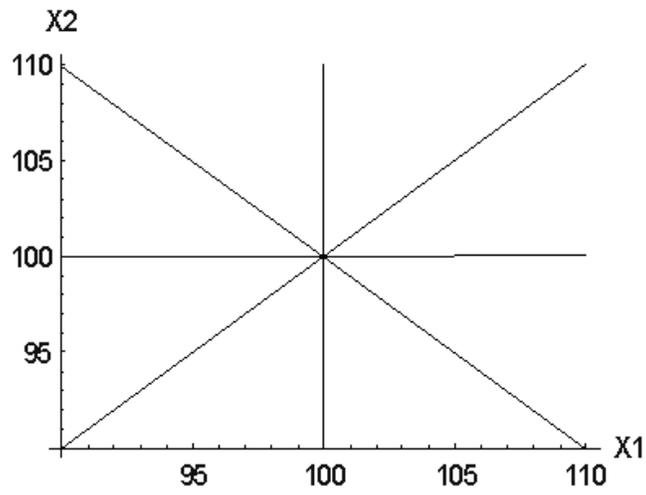